\documentclass[preprint,12pt]{elsarticle}

\usepackage{graphicx}

\journal{Physica E}

\begin{document}

\begin{frontmatter}

\title{Model of circular polarization dependence on Mn delta-layer position in LED heterostructures
       with InGaAs/GaAs quantum well}
\author{D.V. Khomitsky\corref{cor1}}
\ead{khomitsky@phys.unn.ru}
\cortext[cor1]{Tel: +79103836601, Fax: +78314658592 }
\address{Department of Physics, University of Nizhny Novgorod,
         23 Gagarin Avenue, 603950 Nizhny Novgorod, Russian Federation}

\begin{abstract}
A simple model of circular polarization dependence on Mn delta-layer position in LED heterostructures
with InGaAs/GaAs quantum well is proposed, being able to explain quite accurately recent fascinating
experimental results [S.V. Zaitsev, et al., Physica E (2009), doi:10.1016/j.physe.2008.11.003]. The model
emphasizes the role of position-dependent exchange interaction between injected holes and Mn spins
which significantly affects the hole level splitting in a magnetic field, leading to
strong variations of polarization dependence. The role of effective temperature
corresponding to the injected hole energy and the broadening caused by the geometrical
structure imperfections is also discussed.
\end{abstract}


\end{frontmatter}

\section{Introduction}

In recent experiments \cite{Zaitsev,Dorokhin} an intriguing strong enhancement of
polarization degree in LED heterostructures with InGaAs/GaAs quantum well and Mn delta-layer
with varying position has been reported. Since the early 1980-s the properties of circular
polarized emission from semiconductor structures doped with magnetic atoms attract
a considerable attention, and nowadays it became even more important in the scope of nanostructure
design and possible applications in nanophotonics, tunable spin transport and quantum information
processing. Firstly, the dependence of circular polarization on the magnetic field
has been studied both experimentally and theoretically in bulk GaAs samples doped by
Mn acceptors \cite{Karlik,Averkiev}. In these pioneering papers a simple and effective model of
polarization dependence has been derived which later was successfully applied for
the explanation of several recent experiments with bulk samples
\cite{Sapega06,Sapega07}. Later, as the focus of research has moved to the low-dimensional nanostructures,
the triangular quantum wells delta-doped with Mn inside the well have been fabricated
and studied \cite{Nazmul03,Nazmul08,Meilikhov} as well as the structures with
delta Mn layer neighboring to the rectangular quantum well where interesting and promising magnetic
properties of Mn layer have been discovered \cite{Aronzon}. Another important type of
interaction that can affect the polarization and relaxation of spins in quantum wells
is the spin-orbit interaction \cite{Sherman} which is not included in
the present model due to the stronger influence of the Mn d-electrons compared
to the a relatively weak spin-orbit effects in rectangular InGaAs quantum wells.
In the present paper paper we derive a simple quantitative model describing the
magnetic field dependencies of circular polarization in InGaAs/GaAs quantum wells on the position
of Mn delta-layer reported recently. We focus on the role of wavefunction overlap and exchange
interaction between holes and Mn atoms which modify the hole energy spectrum and produce
a measurable effect of the polarization dependence. The model presented here is
intended to clarify only the primary features of the dependencies of polarization curves
on the Mn delta-layer position. Hence, the manuscript is focused on the polarization
dependencies only and it does not touch the fundamental and applied problems
of magnetic and material properties of Mn layer which are of great importance for
condensed matter physics and its applications \cite{Nazmul03,Nazmul08,Aronzon}.
Our paper is organized as follows: in Section 2 we briefly describe the experimental results
obtained in \cite{Zaitsev,Dorokhin} which were the starting point of our studies,
in Section 3 we derive the model of polarization dependence, in Section 4 we discuss the results,
and the conclusions are given in Section 5.

\section{Experimental results obtained by other authors \cite{Zaitsev,Dorokhin}}

\begin{figure}
  \centering
  \includegraphics[width=85mm]{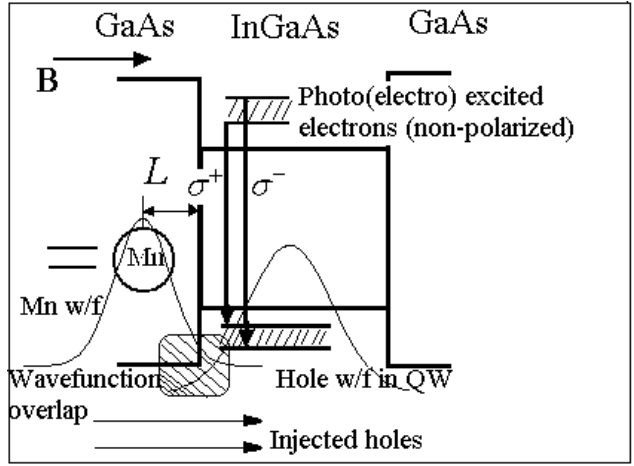}
  \caption{Schematic band diagram, wavefunction overlap and the circular polarized
           transitions in the experiments on Mn delta-doped InGaAs/GaAs QW
           \cite{Zaitsev,Dorokhin}. The holes were injected into the area of 10-nm-thick InGaAs/GaAs
           quantum well, and then the photo- and electroluminescence have been measured in the Faraday
           geometry with the magnetic field perpendicular to the QW plane (pointed by the horizontal
           arrow ${\bf B}$). The downward arrows indicate the transitions from the electron to hole
           levels accompanied by the right- and left-polarized emission of photons.}
  \label{fband}
\end{figure}

Here we shall briefly describe the setup and results of the experiments which we are going
to describe theoretically in the manuscript; for more details, see the original papers
\cite{Zaitsev,Dorokhin}. The schematic view of the band structure which emission properties have been
studied in the experiments is presented in Fig.\ref{fband} The holes have been injected into the area of
10-nm-thick InGaAs/GaAs quantum well, and then the photo- and electroluminescence have been measured
in the Faraday geometry with the magnetic field perpendicular to the QW plane (pointed by the horizontal
arrow ${\bf B}$). The downward arrows indicate the transitions from the electron to hole levels accompanied by
the right- and left-polarized emission of photons. The Mn delta-layer position located on the distance $L$
from the interface is shown schematically in the left part of GaAs layer, and the overlap of Mn and
hole wavefunctions is indicated in the artificially magnified scope just for
the qualitative explanation of the geometrical layout.

\begin{figure}
  \centering
  \includegraphics[width=85mm]{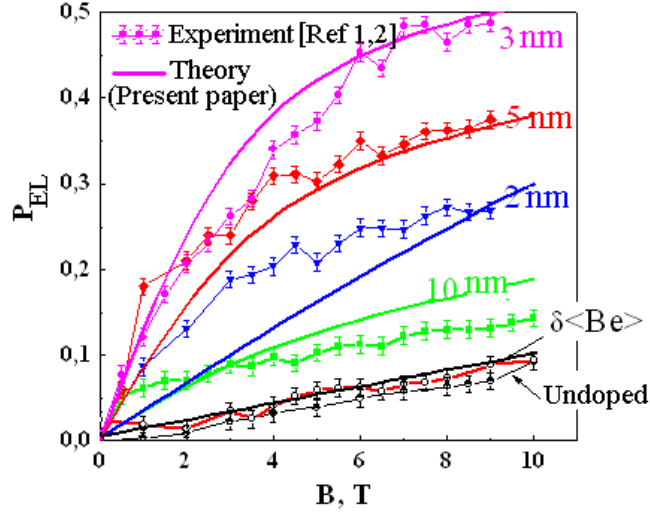}
  \caption{(Color online) Comparison of experimental \cite{Zaitsev,Dorokhin} observations
           (vertically dashed lines) and the presently derived model (solid lines).
            Various curves correspond to different values of distance $L$ from the Mn layer
           to the interface, and two close lowest curves correspond to the reference (undoped)
           and Be-doped (nonmagnetic) structure.}
  \label{fdep}
\end{figure}

The experimental results together with the theoretical calculations presented below
are shown in Fig.\ref{fdep}. The experimental polarization dependencies (vertically dashed lines)
are plotted by using the results of the original paper \cite{Zaitsev}, and the results of our model are
shown as solid lines. Various curves correspond to different values of distance $L$ from the Mn layer
to the interface, and two close lowest curves correspond to the reference (undoped) and Be-doped (nonmagnetic)
structure. One can see that the model provides a good approximation to
the experimental results except the low-field regime for $L=2$ nm. These properties
and their convergence to the measured results will be described in details below.

\section{Model for calculations}

We start the description of the polarization dependence on the applied magnetic field
with the Hamiltonian of holes in QW which includes the orbital part $H_0$, the Zeeman
part $H_Z=m_J g_h \mu_B B$, and the exchange part $H_{ex}=- A \left({\bf J} \cdot {\bf S} \right)$
which arises due to the presence of Mn delta-layer in the close proximity to the QW:

\begin{equation}
H=H_0+ m_J g_h \mu_B B - A \left({\bf J} \cdot {\bf S} \right)
\label{ham}
\end{equation}

where $\mu_B$ is the Bohr magneton, $m_J$ and $g_h$ are the hole angular quantum number and g-factor,
respectively, $B$ is the magnetic field amplitude, and the exchange operator is described by
the coupling parameter $A$ and the scalar product of the hole spin operator ${\bf J}$ and
the Mn d-electrons spin operator ${\bf S}$ \cite{Averkiev}. The exchange coupling parameter $A=A(L)$
in our model is an ensemble-average quantity and its dependence on the delta-layer distance $L$ from
the QW edge will be discussed in details below.

First, we describe briefly the origin of the strong dependence of polarization on the Mn delta-layer
position. The Zeeman term corresponds to the energy $E_Z(B)=\pm \frac{3}{2} g_h \mu_B B$ for two
lowest hole levels with $m_J = \pm 3/2$ where $g_h \approx 3$ for holes in the strained
InGaAs/GaAs structure \cite{Lin}. Even in strong magnetic fields up to $B=10$ T
the Zeeman term provides only about $2$ meV in energy splitting. Below we shall see that
in addition to the Zeeman term the exchange coupling (the last term in Eq.(\ref{ham}))
produces a considerable effect in the hole level splitting and, hence, in
the polarization dependence. Indeed, it is known that the coupling constant $A$ in
GaAs is of the order of $4$ meV for the case of Mn atoms located in the same region
with holes in bulk sample \cite{Averkiev}. If the angular quantum number $m_J= \pm 3/2$
for holes and $m_d = \pm 5/2$ for the Mn d-electrons, than the contribution from
the exchange term in (\ref{ham}) can reach as high as $15$ meV which is almost an
order of magnitude greater than the Zeeman energy alone. We believe that such considerable
effect of exchange coupling can be an origin of strong enhancement of polarization
degree and its sensitive dependence on the Mn delta-layer position. The precise value
of the hole g-factor can not make a big difference here since it only enters the
Zeeman term which, as we see, can be weaker than the exchange term. Our model shows
that the value $g_h = 3$ is in the agreement with the experimental data from the
reference (undoped) sample (lowest curve labeled as "Undoped" in Fig.\ref{fdep}).

We now turn to the detailed description of the theoretical model.
The energy shift caused by the exchange interaction can be described by the average
spin ${\bar S_d}(B,T)$ of the Mn d-electrons which interact with hole spins
$m_J = \pm 3/2$:

\begin{equation}
E_{ex}(B,T,L) = \pm \frac{3}{2} {\bar S_d}(B,T) A(L).
\label{enex}
\end{equation}

The arguments in Eq.(\ref{enex}) mean that the average spin of d-electrons in Mn depends
both on the magnetic field and on the temperature $T$ which takes into account both thermal
and structural level broadening. We do not touch here the problem of possible ferro- or
antiferromagnetic properties of Mn delta-layers in semiconductor heterostructures
\cite{Aronzon} since we are interested in the averaged macroscopic properties of
the delta-layer rather than in its micro- or mesoscopic magnetic structure.
Since the Mn concentration was $0.4$ of the monolayer \cite{Zaitsev}, its statistical
energy distribution function can be taken of the Boltzman form.
Hence, we model the ensemble averaged spin of Mn atoms subjected to magnetic field as
\cite{LL5}

\begin{equation}
{\bar S_d}(B,T)=\frac{1}{Z} \sum_{m_d}
\exp \left( -\frac{m_d g_e \mu_B B}{T} \right)
\label{sdaver}
\end{equation}

where $Z$ is the statistical sum and the Mn d-electrons g-factor $g_e=2$
\cite{Karlik,Averkiev}. In Eq.(\ref{sdaver}) the Mn d-electron spin projections
$m_d=-5/2,-3/2, \ldots, 3/2,5/2$.

The next step is the calculation of the dependence of exchange coupling parameter $A(L)$
on the distance $L$ between the Mn delta-layer and the heterostructure interface.
Let us start with a simple result for the exchange coupling constant for two spherical
atoms with energies $E_{1,2}$ which depends on the inter-atomic spacing  shift $L-L_0$ as \cite{LL3}

\begin{equation}
A(L) \propto A(L_0) e^{-(\kappa_1+\kappa_2)\mid L-L_0 \mid}
\label{al}
\end{equation}

where $\kappa_{1,2}=\sqrt{2m \mid E_{1,2} \mid}/\hbar$. We believe that this simple
result can be applied also in our case as a starting point since both the holes and
the Mn atoms are characterized by the well-defined spatial locations in the narrow QW
and in the delta-layer, respectively. According to the experimental data \cite{Zaitsev},
the maximum value of the coupling constant $A(L_0)=4$ meV \cite{Averkiev} when
the delta-layer is located at $L=L_0=3$ nm (see Fig.\ref{fdep}).
By comparing the acceptor Bohr radius in GaAs which is of the order of $2$ nm together
with the fluctuations in the individual delta-layer atomic positions with respect to
the InGaAs/GaAs interface, the estimated barrier penetration length of the hole wavefunction
may reach up to  $1.5 - 2.0$ nm. Thus, one can conclude that in all cases for $L=2,3,5,10$ nm
presented in Fig.\ref{fdep} the Mn atoms will affect the hole energy levels in the QW
via the exchange interaction with typical length in (\ref{al}) equal to
$1/(\kappa_1+\kappa_2) \approx 4$ nm.

Now the exchange energy (\ref{enex}) is determined and we can calculate
the polarization degree $P(B,L)$ by applying the standard definition of
$P=(I_{+}-I_{-})/(I_{+}+I_{-})$ where $I_{\pm}$ is the right- and left-polarized
radiation intensity, respectively and by using the relations for $P$
arising from the comparison of the level population \cite{Karlik,Averkiev}, which leads to
the following result of the well-known form:

\begin{equation}
P(B,L)=\tanh \left[ x(B,L) \right]
\label{pdep}
\end{equation}

where the argument

\begin{equation}
x(B,L)=\frac{E_Z(B)+E_{ex}(B,T,L)}{T_{eff}}
\label{x}
\end{equation}

depends not only on the reservoir temperature or the fluctuation field amplitude (whichever is greater)
$T\sim 7$ K \cite{Karlik,Averkiev} but also on the effective temperature $T_{eff}$ which may be different
from $T$. The meaning of the effective temperature here reflects the fact that
the incoming injected holes participating in the recombination process are
non-equilibrium hot holes, so they can not be described by the reservoir temperature $T$.
Of course, the detailed properties of such hot holes including their energy
distribution deserve a separate investigation but for our purposes the primary
relevant quantity is their average energy (or temperature). The value of $T_{eff}$ can be
estimated from the voltage drop since the holes gain the kinetic energy from this
source and the resulting value can be tested by applying Eq.(\ref{pdep}) to the
reference undoped sample. Following this approach, we find that $T_{eff} \approx 20$ meV
which is in agreement with the real experimental setup where the total voltage drop of
$1\ldots 2$ V has been applied to the structure of $0.5$ microns of total thickness.

\section{Results and discussion}

The calculated polarization dependence (\ref{pdep}) corresponding to various positions of the Mn
delta-layer is presented in Fig.\ref{fdep} together with the experimental results previously obtained
by other authors \cite{Zaitsev,Dorokhin}. Various curves correspond to different values of distance $L$ from
the Mn layer to the interface, and two close lowest curves correspond to the reference (undoped)
and Be-doped (nonmagnetic) structure. One can see that the model derived here provides
a satisfactory approximation of the experimental observations everywhere except the low-field regime
for the closest position of the Mn layer at $L=2$ nm and the high-field area at $L=10$ nm which
is discussed below. Indeed, the effects from the exchange interaction can lead to a substantial increase
of the polarization degree which rises when the Mn delta-layer is located closer to the QW and
its interaction with holes is increased. This dependence on $L$ holds until the region of
very close proximity $L=2$ nm is reached where the amplitude of the polarization degree
drops and the theoretical calculations diverge from the experimental data in
the region of low magnetic field. We believe that such discrepence is coming from
the increasing microscopic geometrical disorder in the position of Mn atoms as long
as the delta-layer is moved very close to the interface. This assumption is supported
by the meaning of the temperature $T$ in Eq.(\ref{x}) which takes into account not
only the reservoir temperature but also the level broadening caused by the defects,
impurities and the fluctuating field which role has been recognized in the earliest
studies of this problem \cite{Karlik,Averkiev}. Our calculations have shown that
even at the reservoir temperature $T_{0}=2$ K the actual value of $T$ to reproduce
the experimental data is close to $7$ K which is in the agreement with level
broadening of $0.6$ meV even in high purity samples. It is known that a fluctuation of
the QW width of just one or two atomic monolayers can lead to the broadening of the optically
detected emission lines of the order of $15$ meV \cite{Ganichev}. In the experiments with Mn doped
GaAs structures the amplitudes of the fluctuating field were of the order of $1$ meV corresponding to
$T \approx 11$ K. Hence, when the geometrical disorder is increased at $L=2$ nm, some higher values
of $T$ are expected, and we found that the actual level broadening here is satisfactory described
by $4.5$ meV corresponding to $T=50$ K. We stress that this deviation from the previously obtained
uniform set of parameters ($T=7$ K, $T_{eff}=20$ meV) is the only one which we allowed through
the calculations and we believe that it can be adequately explained by the increase of the geometrical
disorder caused by the very close proximity of the doping layer to the QW interface.
The slight deviation between theory and experiment at high fields and $L=10$ nm (see Fig.\ref{fdep})
can be explained by the coupling position dependence between Mn atoms and the holes
which can decrease a little faster at high distances for the isolated Mn delta-layer and two-dimensional
holes confined in the QW than the simple approximation (\ref{al}) predicts.

\section{Conclusions}

We have derived a quantitative model of circular polarization dependence on Mn delta-layer
position in LED heterostructures with InGaAs/GaAs quantum well which is able to explain quite accurately
the experimental results. We have studied the role of position-dependent exchange interaction between
injected holes and spins of Mn d-electrons which significantly affects the hole level splitting in
a magnetic field, leading to strong variations of polarization dependence as a function of the delta-layer
position. The injected hole energy and the level broadening caused by the geometrical structure disorder
are also included in the model providing a satisfactory level of convergence between experimental observations
and theoretical calculations.

\section*{Acknowledgements}
The author is grateful to Yu.A. Danilov, V.D. Kulakovskii, S.V. Zaitsev and M.V. Dorokhin
for useful discussions. The work has been supported by the RNP Program of Ministry of Education
and Science RF, by the CRDF, and by the RFBR.

\end{document}